\documentclass[amsmath, amssymb,11pt]{article}
\usepackage{moriond,epsfig}
\usepackage{bm}
\usepackage{amsfonts}
\newcommand{\half}[0]{\frac{1}{2}}

\newcommand{\ld}[0]{\mathcal{L}}

\newcommand{\dd}[0]{\textrm{d}}

\newcommand{\gbm}[1]{\bm{#1}}

\def\Journal#1#2#3#4{{#1} {\bf #2}, #3 (#4)}

\def\PLB{{\em Phys. Lett.}  B}

\def\PRD{{\em Phys. Rev.} D}

\def\JCAP{{\em JCAP} }

\def\be{\begin{equation}}
\def\ee{\end{equation}}
\def\bea{\begin{eqnarray}}
\def\eea{\end{eqnarray}}

\begin{document}
\title{Charge, domain walls and dark energy}

\author{JONATHAN A PEARSON}

\address{Jodrell Bank Centre for Astrophysics, School of Physics and Astronomy, \\The University of Manchester, \\Manchester M13 9PL, U.K\\15/10/10}

\maketitle\abstracts{
One idea to explain the mysterious dark energy which appears to pervade the Universe is that it is due to a network of domain walls which has frozen into some kind of static configuration, akin to a soap film. Such models predict an equation of state with $w=P/\rho=-2/3$ and can be represented in cosmological perturbation theory by an elastic medium with rigidity and a relativistic sound speed. An important question is whether such a network can be created from random initial conditions. We consider various models which allow the formation of domain walls, and present results from an extensive set of numerical investigations. The idea is to give a mechanism which prevents the natural propensity of domain walls to collapse and lose energy, almost to the point where a domain wall network freezes in. We show that when domain walls couple to a field with a conserved charge, there is a parameter range for which charge condenses onto the domain walls, providing a resistive force to the otherwise natural collapse of the walls.}

\section{Introduction and motivation}
Modern cosmology has found itself a problem. If observational data are understood within the framework of the FRW model in General Relativity, one must invent a  ``dark energy'' component within the content of the universe, to explain acceleration (in the absence of a cosmological constant). The alternative is to move away from the FRW model, or to modify the gravitational theory.

In this paper we provide fresh theoretical evidence for the viability of the elastic dark energy model -- a model which uses frozen domain wall networks to accelerate the universe~\cite{ede}. The equation of state for a gas of domain walls moving at velocity $v$ is $w_{{\rm dw}} = v^2 -{2}/{3}$, hence our requirement for a frustrated ``frozen'' network, for which $w_{\rm dw} = -2/3$. We present a mechanism which freezes domain wall networks. If symmetry currents interact with domain walls, a  pseudo-stable glass-like network forms. 

We will begin by giving a quick review of features and results of the proto-typical domain wall forming model. We will then discuss two models, which have the feature that a field with discrete minima interacts with a second field with continuous $U(1)$ symmetry.

\section{$\mathbb{Z}_2$ domain walls}
If a field theory has a potential, and if the potential has multiple disconnected vacua, domain walls form. Domain walls are the interpolations between spatially adjacent vacuum states~\cite{V&S}. The simplest example of such a field theory is described by the Lagrangian
\bea
\label{jp:eq:z2_lag}
\ld = \half \partial_{\mu}\phi \partial^{\mu}\phi - \frac{\lambda}{4}\left( \phi^2 - \eta^2\right)^2.
\eea
The vacuum in this theory -- the configurations of the scalar field $\phi(x^{\mu})\in\mathbb{R}$ that minimise the potential -- is just the set $\mathcal{V} = \left\{ +\eta, -\eta\right\}$. If we imagine two chunks of space, one  with $\phi = +\eta$ and another with $\phi = -\eta$, then the field must continuously interpolate at the boundary between the ``domains''. There is a static solution to the equations of motion which captures such an interpolation, $\phi(x) = \eta_{\phi} \tanh(x/\Delta)$, where $\Delta := \sqrt{2/\lambda \eta^2}$, the so-called kink solution. The symmetry properties of this model are rather interesting. Consider the full model (\ref{jp:eq:z2_lag}), and transforming $\phi \rightarrow -\phi$: the model is totally invariant under this transformation. However, if one expands about each of the minima, the symmetry is lost. This is a simple example of spontaneous symmetry breaking.

By a rather simple dimensional argument, as well as substantial numerical investigations~\cite{BattyeMoss}~\cite{Avelino_etal}, the number of domain walls formed by the standard Kibble scenario drops, $n\propto t^{-1}$. This result is also observed in many other models (so long as discrete vacua are present) -- even with 3 or more vacua (with junctions). The standard result, therefore, is that domain walls collapse under their own tension.

\textit{\textbf{Numerical simulation}} Throughout the rest of the paper we will refer to simulations that we have performed. We evolve the equations of motion in $(2+1)$-dimensions, using a leapfrog evolver, imposing periodic boundary conditions. We use random initial conditions: each grid-point is randomly assigned to be in one of the vacuum states of the theory. We also apply damping for a small amount of the total simulation time, to ease condensation into domains: results are only meaningful once this damped period has elapsed.

\section{Kinky vortons}
The kinky vorton model allows a charged condensate to live on a domain wall kink solution. The model has Lagrangian
\bea
\ld = \partial_{\mu}\phi \partial^{\mu}\phi + \partial_{\mu}\sigma \partial^{\mu}\bar{\sigma}- \frac{\lambda_{\phi}}{4}\left( \phi^2 - \eta_{\phi}^2 \right)^2 - \frac{\lambda_{\sigma}}{4}\left( |\sigma|^2 - \eta_{\sigma}^2 \right)^2 - \beta \phi^2 |\sigma|^2.
\eea
The first two terms are the kinetic terms of the real and complex scalar fields, $\phi, \sigma$, respectively. Third is the symmetry breaking term of the real scalar -- this allows domain walls to form. Next is a term which has a $U(1)$ symmetry, where the complex scalar takes on a non-zero value in the vacuum. Finally, a quartic interaction term between the fields. This theory has a global $\mathbb{Z}_2 \times U(1)$ symmetry. The model parameters are chosen such that the $\mathbb{Z}_2$ symmetry is broken in the vacuum, and the $U(1)$ symmetry retained. A consequence of the $U(1)$ symmetry is that there exists a conserved symmetry current and charge (Noether's theorem):
\[
J_{\mu} := \bar{\sigma}\partial_{\mu}\sigma - \sigma \partial_{\mu} \bar{\sigma},\qquad \partial_{\mu}J^{\mu}=0, \qquad  \frac{\dd }{\dd t} \int \dd^3x \,J_0=0.
\]

One will immediately notice that there are a large number of model parameters, whose values must be chosen. A range of different choices have been discussed in ref.~\cite{JAP_RAB_charge}, but in the present paper will only consider a single parameter set: $\lambda_{\phi} = \lambda_{\sigma} =2, \beta =\eta_{\phi}=1, \eta_{\sigma} = \sqrt{3}/2$.

A kinky vorton is a stable solution to the Euler-Lagrange equations of motion, which corresponds to a circular kink solution with a current-carrying charged condensate living on the domain wall~\cite{BattyeSutcliffe_KV}. The radius of the loop is entirely determined by the charge and current that reside on the loop: one can imagine the loop being stabilized by the existence of the conserved charge/current (without the condensate, the loop collapses).

\subsection{Formation from random initial conditions}
The loops just described are entirely idealized: single loops, carefully setup in numerical simulations. So, we ask the question: can these charged loops form from random initial conditions, and can they aid network stabilization? 

\begin {figure}[!ht]
      \begin{center}
		{\includegraphics[scale=0.425]{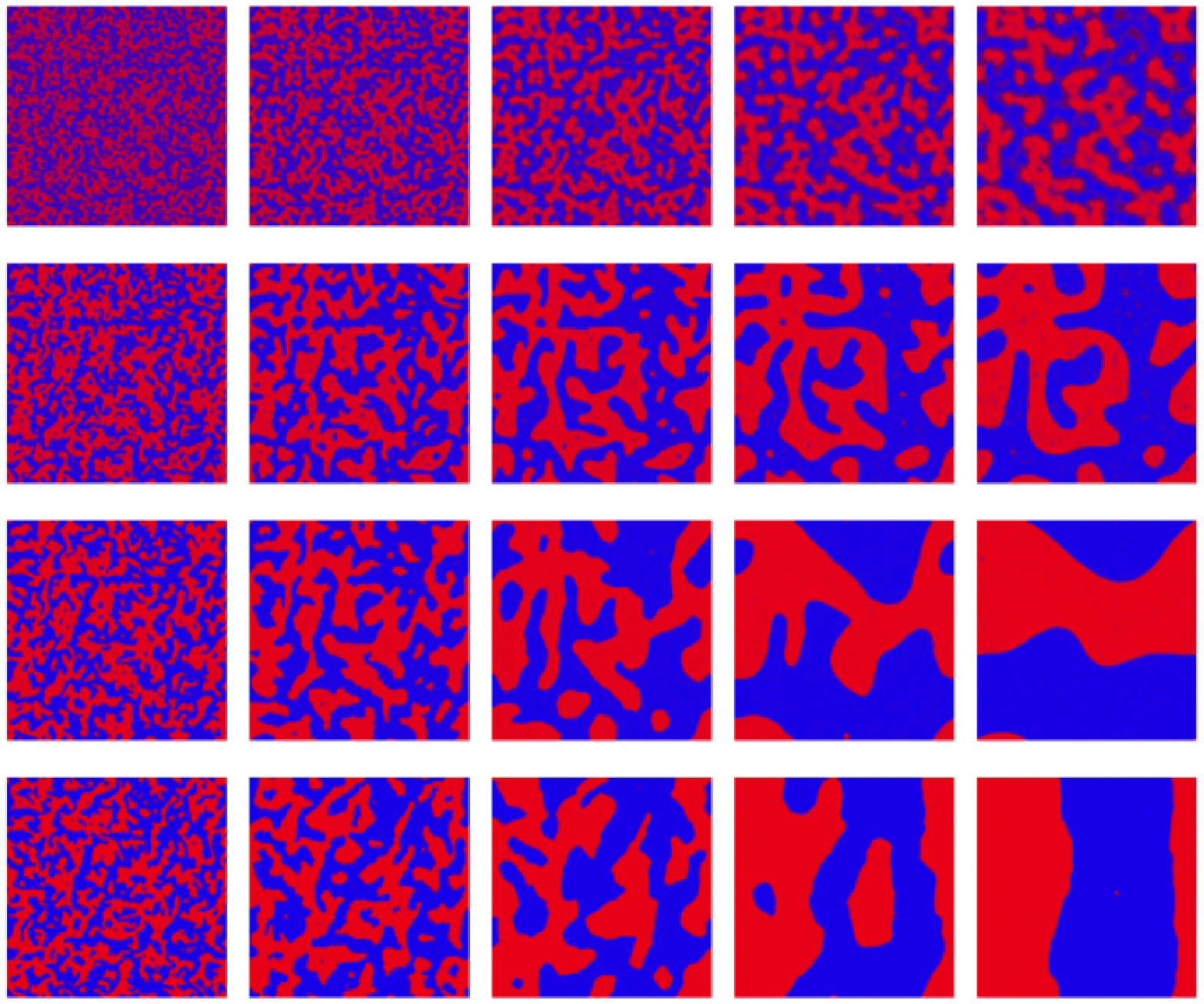}}\qquad
		{\includegraphics[scale=0.517]{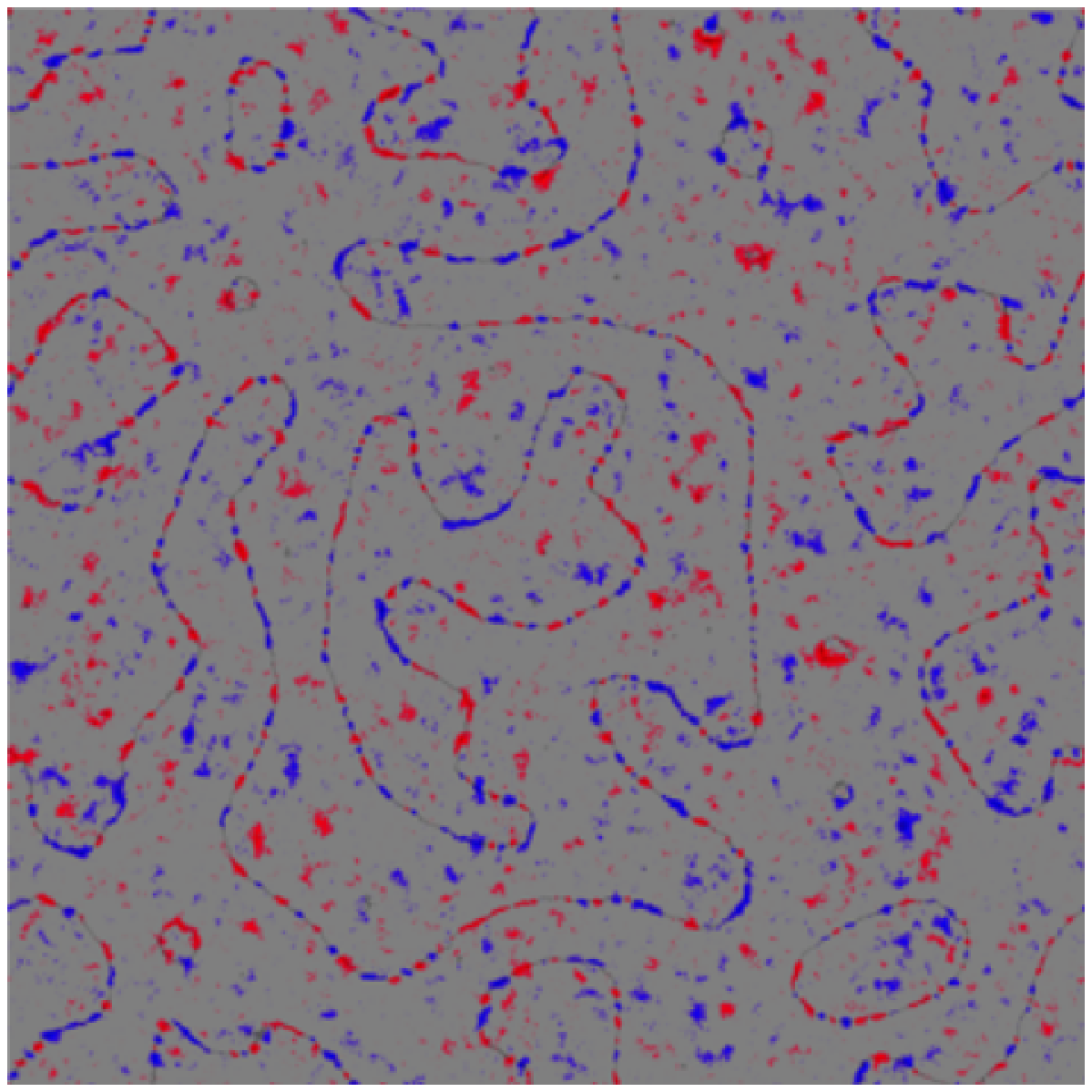}}
      \end{center}
\caption{\textit{Left} Images of the evolution of the $\phi$-field at $t = 80, 160, 320, 640, 1280$ (left to right) for initial charge densities $\rho_{\rm Q}(0) = 0, 0.01, 0.09, 0.25$ (bottom to top). \textit{Right} Image of $\Re(\sigma)$, of the $\rho_{\rm Q}(0) = 0.09$ simulation at $t = 640$. In both sets of images, red/blue denotes positive/negative values (i.e. the two vacua, and a winding, respectively).}\label{jp:fig:kv_imgs}
\end {figure}

We initially setup a homogeneous charged background, and random vacuum occupation. We specify an initial charge density $\rho_{\rm Q}(0)$, and let the equations of motion evolve. Figure \ref{jp:fig:kv_imgs} has images of the $\phi$-field, over time, for various initial charge densities $\rho_{\rm Q}(0)$. It is clear from inspecting these images that as the initial charge is increased, the rate at which the network ``dilutes'' slows down. This is quantified by inspecting a plot of the evolution of the total number of domain walls, for various initial charge densities (not included in the present paper, but can be found another publication~\cite{KVFORM}). We have also isolated some of the loops, and analyzed their properties (winding number, charge and radius). We have shown that they have properties very close to kinky vortons.

\section{Charge-coupled cubic anisotropy model}
The charged-coupled cubic anisotropy model is constructed by interacting a real vector field $\gbm{\phi}(x^{\mu})\in \mathbb{R}^N$ with a complex scalar field $\sigma$, so that the model has a broken $O(N)\times U(1)$ global symmetry. It is only the $O(N)$ symmetry that is broken in the vacuum -- so that domain walls form, and there exists a conserved charge. The potential is
\begin{eqnarray}
V = \frac{\lambda_{\phi}}{4}\left( |\gbm{\phi}|^2 - \eta_{\phi}^2\right)^2+ \epsilon \sum_{i=1}^N\phi_i^4+
 \frac{\lambda_{\sigma}}{4}\left( |\sigma|^2 - \eta_{\sigma}^2 \right)^2 +\beta|\gbm{\phi}|^2|\sigma|^2.
\end{eqnarray}
The first two terms create discrete vacua in the vector field, then the introduction of the condensate field, and finally the term which interacts the condensate and vector fields. As there are more than 2 vacua, junctions and multiple tensions of domain walls can form (a junction is where more than two vacua cycle around a point in space: if four vacua cycle, an $X$-type junction is formed, and if there are three then a $Y$-type junction). A known result is that the number of domain walls in the cubic anisotropy model drops $n \propto t^{-1}$: junctions do nothing to prevent the collapse of a network~\cite{BattyeMoss}. 

We have performed a set of simulations with this model, taking $\gbm{\phi}(x^{\mu})\in \mathbb{R}^2$ . As in the kinky vorton model, we start off with homogeneous initial charge and random initial domain occupation (this time there are 4 vacua/colours to choose from). The parameter choice in this model becomes less trivial, but an interesting parameter worthy of note is the interaction constant $\beta$, where we take $4\beta^2 = \lambda_{\phi} \lambda_{\sigma} + 4\epsilon \lambda_{\sigma}$, chosen to avoid the phase-separation phenomenon~\cite{JAP_RAB_charge}.

\begin {figure}[!ht]
      \begin{center}
		{\includegraphics[scale=0.6]{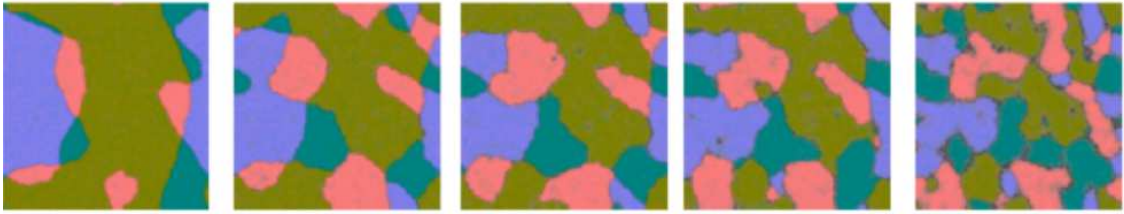}}
      \end{center}
\caption{Images of the $\gbm{\phi}$-field at $t=320$, for $\rho_{\rm Q}(0) = 0.09, 0.16, 0.2, 0.25, 0.36$.}\label{fig:4096-nwalls-varyalpha}
\end {figure}

In Figure \ref{fig:4096-nwalls-varyalpha} we display images of the $\gbm{\phi}$-field, all of which are at the same time-step, but with different initial charge densities. Again, the implication is clear: more charge freezes the pattern in more. Plots of $\Re(\sigma)$, charge and current densities look qualitatively identical to their kinky vorton counterparts, having become associated exclusively with domain walls.

\section{Conclusions and summary}
Our first result is that networks with higher initial charge density freezes wall networks in ``more''.
Secondly, charge and current condense on domain walls, and almost exclusively live on them.  Finally, some of the loops that form have the specific properties of kinky vortons.

The elastic dark energy model was proposed to use domain walls as a model, requiring the wall network to freeze in. The conserved charge of the condensate field discussed here provides a resisting force against the tension of a domain wall, enabling a network to become pseudo-stable.

\section*{Acknowledgements}
Proceedings of the 45th Rencontres de Moriond, cosmology conference.
The work presented here was carried out in collaboration with R. Battye, P. Sutcliffe and S. Pike; our image creation software was largely developed by C. Welshman.

\end{document}